\documentstyle[preprint,aps]{revtex}

\begin{document}
\draft

\preprint{\begin{tabular}{l}
\hbox to\hsize{September, 1994 \hfill SNUTP 94-45}\\[-3mm]
\hbox to\hsize{hep-ph/9409420}\\[5mm] \end{tabular} }

\bigskip

\bigskip

\title{\bf Cosmological Constant
and Supergravity Soft Terms\thanks{Talk presented
at the ITP Workshop
on ``Frontiers in QFT", Beijing, June 15--24, 1994.} }
\author{Jihn E. Kim}

\address{\sl
Department of Physics and Center for Theoretical Physics,
Seoul National University\\
Seoul 151-742, Korea}
\def\be{\begin{equation}}
\def\ee{\end{equation}}
\baselineskip=12pt

\maketitle
\begin{abstract}
In supergravity models, the quantum correction to the vacuum energy
can be of order $M_I^4$, if the cutoff is of order the Planck
mass $M^2_{P}$ and $Str\ {\cal M}^2\ne 0$.  Therefore, the tree level
cosmological constant must be nonzero (probably negative).
Since the current practice of calculating soft parameters in supergravity
models assumes the vanishing tree level cosmological constant, the
supergravity calculation must be accordingly modified.
This implies that the soft parameters in supergravity has an
additional contribution depending on the nonvanishing
tree level cosmological constant. A simple quantum mechanical model
mimicking this situation is also presented.
\end{abstract}


There are three well-known hierarchy problems: the cosmological constant
problem, the Kaluza--Klein problem, and the gauge hierarchy problem.  If
the theoretical physics in 1917 were as advanced as now, the introduction
of the cosmological constant \cite{einstein}
might have had led to the immediate theoretical problem,
the cosmological constant problem.  In the same vein, the Kaluza--Klein
theory \cite{klein} introduced a huge ratio,
$M_{P}/m_e\gg 1$, working against the original beautiful idea, which
is the reason that the Kaluza--Klein theory seems currently
to be in disfavor.  The
third problem is the well-known gauge hierarchy problem in grand unified
theories \cite{gildener}.  In this talk I discuss the cosmological constant
in supergravity, and hope that it instigate young Chinese physicists to
think about this eighty years old problem again and get an idea toward the
solution.  Certainly my talk is not dealing with a solution, but can give
an idea toward a correct direction of the solution.

This talk is based on my recent work collaborated with
Kiwoon Choi and Hans Peter Nilles on the cosmological constant in
supergravity and its effect to soft terms \cite{ckn}.

Supersymmetry (SUSY)  is widely believed to be a
leading candidate for a further symmetry beyond the standard model.
This is largely due  to the fact that SUSY provides
the only known perturbative solution to the problem of quadratic
divergence in the Higgs boson mass.
The most popular working scenario is supergravity models in which
SUSY is broken at the intermediate scale \cite{nilles}.

At low energy, the effective theory is parametrized by SUSY breaking
soft parameters in addition to the renormalizable couplings allowed
by the gauge symmetry. An interesting feature of supergravity models
is that  many of the coefficients of soft terms are
calculable at tree approximation.
In some cases, one obtains certain tree level relations
among the soft coefficients renormalized at the Planck scale $M_P$.
For instance, one obtains $A=-\sqrt{3}m_0\lambda$ in the
dilaton-dominated SUSY breaking scenario in string theory
$\cite{louis,ibanez}$,
where $A$ denotes a generic trilinear scalar coefficient,
$\lambda$ is the associated Yukawa coupling, and $m_0$
is the soft scalar mass.

For physical applications of such tree level results,
one has to  take into account quantum corrections.
Ordinary renormalizable interactions lead to
one loop corrections proportional to
\begin{equation}
\frac{g^2}{8\pi^2}\ln(\Lambda^2/\mu^2)
\end{equation}
where  $g$ denotes a gauge or Yukawa coupling
constant,  $\mu$ is a scale around the weak scale $m_W$
where the loop graph is being evaluated,
and $\Lambda$ is the momentum cutoff above which the validity
of four-dimensional $N=1$ supergravity theory breaks down.
Clearly for $g^2$ of order unity and $\Lambda$ not far below $M_P$,
the corrections become important due to a large logarithm.
A nice feature of this type of corrections  is that they depend
on the cutoff $\Lambda$ logarithmically and are  {\it calculable} within
the supergravity model by the aid of renormalization group analysis.

Besides the above type of calculable corrections, there
are other types of corrections which depend strongly on $\Lambda$ and
thus whose precise magnitudes are {\it not} calculable
 within the supergravity model $\cite{gaume}$.
For instance, one loop graphs induced by nonrenormalizable gravitational
interactions would give power-law divergent corrections
which are proportional to
\begin{equation}
\frac{1}{8\pi^2}(\kappa\Lambda)^n,
\end{equation}
where $\kappa=\sqrt{8\pi}/M_P$ denotes the dimensionful
gravitational coupling constant and $n$ is a positive integer.
Obviously, the main correction in this case arises from
the fluctuations at the scale $\Lambda$.
As a result, they are strongly dependent on the unknown physics at
the scale $\Lambda$.  Thus one has to know the underlying theory
in more detail to get a useful information on such parameters.

Roughly speaking, the dimensionless coupling `$\kappa\Lambda$'
measures how strongly the underlying theory of supergravity
is coupled.  If strongly coupled, one can guess that
$\kappa\Lambda\simeq 4\pi$ which means that loops are as important
as trees.  (This is analogous to
the nonlinear sigma model of the pions whose underlying theory
is the strongly coupled QCD$\cite{georgi}$.)
In this case, tree level predictions would not be useful
at all {\it unless} they are  mere consequence of some symmetries
of the underlying theory.

Of course, one can ignore the incalculable power-law divergent
corrections if the underlying theory is very weakly coupled, i.e.
$\kappa\Lambda\ll 1$.  However, a more reasonable scenario would
be that the underlying theory is moderately weakly coupled with
$\kappa\Lambda$ of order unity, giving a loop factor of order
$1/8\pi^2$.  This is indeed the case for supergravity models
which correspond to the low energy limit of  string theory.
In this case, it is natural to set $\Lambda$ to the string scale
\footnote{It has been argued $\cite{dine}$ that another candidate
for the cutoff, the compactification scale, is
quite close to  $M_{\rm string}$.} $M_{\rm string}$.
In the heterotic string theory, the gravitational coupling
constant $\kappa$ is given by
 $\kappa M_{\rm string}=g_{\rm GUT}=O(1)$\cite{dine}
where $g_{\rm GUT}$ is the unified gauge coupling constant
at $M_{\rm string}$.
In view of the string theory, thus the interesting case
is a moderately weakely coupled supergravity, leading to
$\kappa\Lambda$ being of order unity.

For $\kappa\Lambda=g_{\rm GUT}$, the power-law divergent
corrections of Eq. (2) are expected to be  of order
${\alpha_{\rm GUT}}/{2\pi}$ and thus are not so significant.
This seems to be a small correction.  But if there are
sufficiently large number of particles in the loop, then
the correction can be of O(1).  Without supersymmetry, however,
it is difficult to take this effect into account.

With softly broken supersymmetry, the radiative correction to
the coefficients of operators {\bf 1},
$\phi^*\phi$ and $\bar\psi\psi\phi$ are quadratically divergent,
logarithmically divergent and finite, respectively.  The
coefficient of {\bf 1} is the cosmological constant.
The cosmological constant is an important unresolved problem
in spontaneously broken gauge theories \cite{veltman}.
At one loop order, the cosmological constant which is the
vacuum energy density in our notation  receives
a quadratically-divergent zero point energy contribution
$\cite{zumino}$.
For $N$ chiral multiplets with $m_B\gg m_F$,
the contribution is given by
\be
N\int {d^3k\over (2\pi)^3}\left(\sqrt{{\bf k}^2+m_B^2}-\sqrt{{\bf k}^2
+m_F^2}\right) \, \simeq \, N\frac{m_B^2\Lambda^2}{8\pi^2}.
\ee
This leads to corrections
of order $N{\alpha_{\rm GUT}}/{2\pi}$ to some of soft coefficients.
A key point in this regard is that although
$\alpha_{\rm GUT}/2\pi$ is small, $N \alpha_{\rm GUT}/{2\pi}$
can be significantly large since
in realistic models $N$ is typically of order
$8\pi^2$. \footnote{In the minimal supersymmetric standard model
$N=49$, but it can be larger if there are more chiral multiplets
at the intermediate scale.}
Therefore, the cosmological constant is shifted by an amount
of order $M_{I}^4$ which is a natural order of the tree level
cosmological constant.  Therefore, calculations of soft parameters
assuming a vanishing tree level cosmological constant can be
significantly affected.

To be definite, let us consider a simple supergravity model with the
following K$\ddot {\rm a}$hler potential and
superpotential\cite{nilles}
\begin{eqnarray}
K&=&h_{\alpha}h^*_{\alpha}+\phi_i\phi^*_i
+(\xi H_1H_2+h.c.)\ ,
\nonumber \\
W&=&W_h(h_{\alpha})+\frac{1}{6}
\tilde{\lambda}_{ijk}\phi_i\phi_j\phi_k+\tilde{\mu}H_1H_2,
\end{eqnarray}
where $h_{\alpha}$ denote hidden sector
fields triggering SUSY breaking, and
 $\phi_i$ are generic observable sector fields
including quarks, leptons, and the two Higgs doublets $H_1$ and $H_2$.

To obtain the effective action $S_{\phi}$ of the observable fields $\phi_i$,
we integrate out the hidden sector fields to obtain
\be
\exp (i S_{\phi}) = \int [{\cal D} h] \exp (i S),
\ee
where $S$ is the full supergravity action, and
 $[{\cal D} h]$ includes the integration of the gravity
multiplet
$(g_{\mu\nu}, \psi_{\mu})$ over
a background spacetime metric $\bar{g}_{\mu\nu}$ with macroscopic
wavelength. Since the high momentum modes of $\phi_i$ are
not integrated out yet, $S_{\phi}$ is defined at the cutoff
scale $\Lambda$ in the sense of Wilson. Thus to study
the low energy physics of $\phi_i$, one still needs to
scale the renormalization point down to the weak scale.
At any rate,
in the flat limit, $S_{\phi}$ would be characterized by
the effective superpotential of global SUSY$\cite{nilles}$
\be
W_{\rm eff}\ =\ \frac{1}{6}\lambda_{ijk}\phi_i\phi_j\phi_k+\mu H_1H_2,
\ee
and also  the soft breaking part  of the form
\be
{\cal L}_{\rm soft}\ =\ m_0^2\phi_i\phi^*_i
+(\frac{1}{6}A_{ijk}\phi_i\phi_j\phi_k+BH_1H_2+\rm h.c.).
\ee

Among the coefficients in $W_{\rm eff}$ and ${\cal L}_{\rm soft}$,
those which depend
on the expectation value of the hidden sector scalar potential are
relevant for us. They are the soft scalar mass $m_0^2$ and the
$B$ coefficient. To calculate these, we expand the supergravity
potential $V$ in powers of light fields $\phi^*\phi$.
Noting that
\begin{equation}
V\ =\ e^G\left[{\partial G\over \partial z_I}(G^{-1})^J_I{\partial G
\over \partial z^*_J}-3\right]
\end{equation}
where
\begin{equation}
G\ =\ K+\log |W|^2,
\end{equation}
we obtain
\begin{eqnarray}
V\ =\ &&V_h+(m_{3/2}^2+\kappa_0^2V_h)\phi^*_i\phi_i\nonumber \\
&&+[(1+|\xi|^2)m_{3/2}^2+\kappa_0^2V_h](H_1^*H_1+H_2^*H_2) \\
&&+(2m_{3/2}^2+\kappa_0^2V_h)(\xi H_1H_2+\xi^*H_1^*H_2^*)+O(\phi_i^3).
\nonumber \end{eqnarray}
By taking the expectation values of the coefficients of light fields,
the soft parameters are given by
\begin{eqnarray}
m_0^2 &&= \langle m_{3/2}^2+\kappa_0^2V_h\rangle\nonumber \\
m_H^2 &&= \langle (1+|\xi|^2)m_{3/2}^2+\kappa_0^2V_h\rangle \\
B &&= \langle \xi (\kappa_0^2 V_h+2m_{3/2}^2)\rangle .\nonumber
\end{eqnarray}
The other soft parameters and couplings are given by
\begin{eqnarray}
&&\lambda_{ijk}=\langle \tilde{\lambda}_{ijk}\exp (\kappa_0^2
h_{\alpha}h^*_{\alpha}/2)\rangle,\nonumber \\
&&\mu=
\langle (\tilde{\mu}+\xi \kappa_0^2 W_h)
\exp (\kappa_0^2 h_{\alpha} h^*_{\alpha}/2)\rangle , \\
&&A_{ijk}=\langle \tilde{\lambda}_{ijk} \kappa_0^2 (h_{\alpha}
D_{\alpha} W_1)^* \exp (\kappa_0^2 h_{\alpha} h^*_{\alpha})\rangle,
\nonumber
\end{eqnarray}
where $\kappa_0$ is the bare gravitational coupling constant,
 $D_{\alpha}W_h=(\partial_{h_{\alpha}}+\kappa_0^2 h^*_{\alpha})W_h$,
and the gravitino mass $m_{3/2}$ and the hidden sector
scalar potential
$V_h$ are given by
\begin{eqnarray}
&&m_{3/2}^2=\kappa_0^4 |W_h|^2\exp (\kappa_0^2 h_{\alpha} h^*_{\alpha}),
\nonumber \\
&&V_h= (|D_{\alpha} W_h|^2-3\kappa_0^2 |W_h|^2)\exp(\kappa_0^2 h_{\alpha}
h^*_{\alpha}).
\end{eqnarray}
Here
the bracket means the average over the hidden
sector fields.  For example,
\be
\langle V_h\rangle =\int [{\cal D} h] V_h(h_{\alpha}) \exp (iS_h)/
\int [{\cal D} h] \exp (i S_h),
\ee
where $S_h$ is the supergravity action of the hidden sector fields alone.
If interactions among hidden sector fields are weak enough,
which is usually the case,\footnote{
In some cases,
 hidden sector contains gauge interactions
which become strong to provide
a nonperturbative seed, e.g. the gaugino
condensation$\cite{gcondensation}$, for SUSY breaking.
We assume in such cases that
the strongly
interacting gauge nonsinglet sector is already
integrated out, whose effects are included in
the effective
hidden sector superpotential $W_h$ of weakly interacting
gauge singlet $h_{\alpha}$.}, the above discussion applies.
The soft parameters $m_0^2$ and $B$ depends on
$\langle V_h\rangle$.  But $\langle V_h\rangle$ is not
the cosmological constant.

Then, what is the cosmological constant?
The fully renormalized cosmological constant
$V_{\rm eff}$
at low energy is obtained by integrating
out all the fields in the theory,\footnote{The long wave length
metric $\bar g_{\mu\nu}$ is treated as background and is
not integrated out to see the
gravitational effect at low energy.}
\be
\exp (i\int d^4 x V_{\rm eff})=\int [{\cal D}\phi {\cal D}h]
\exp (i\int d^4 x \sqrt{g}{\cal L})
\ee
where $[{\cal D}\phi]$ represents the integration over all
the observable gauge and matter multiplets.

Of course, $\langle V_h\rangle$ and $V_{\rm eff}$ are
different.  At tree level, however, they are the same.
In the classical approximation, $\langle V_h \rangle$
is simply the classical potential
in ${\cal L}_h$ since the path is taken over the fields satisfying
the classical equation of motion. From Eq. (15), the tree level
value of $V_{\rm eff}$ is just the classical potential in
${\cal L}_h$ since the nontrivial
contribution of $[{\cal D}\phi]$ integration is the effects of
loops of $\phi$ fields.  Therefore, at tree level the
cosmological constant $V_{\rm eff}$ and $\langle V_h\rangle$ are the
same.

Note, however,that although $\langle V_h \rangle_{\rm tree}$
corresponds to the {\it tree level} cosmological constant,
the quantity  $\langle V_h\rangle$ that appears in the soft
coefficients of Eq. (11)
is {\it not} the fully renormalized cosmological constant.
Eq. (11) shows that
$m_0^2$ and  the part of $B$
associated with the K\"{a}hler
potential term $(\xi H_1H_2+h.c.)$\cite{masiero}
depend on $\langle V_h\rangle$.
For local SUSY broken by the vacuum value
of the auxiliary component $F_{\alpha}=D_{\alpha} W_h \exp (\kappa_0^2
h_{\alpha} h^*_{\alpha}/2)$, {\it unless} one implements
 fine tuning of
some parameters in the hidden sector
superpotential $W_h$, the typical
size of $\langle V_h\rangle$ would be of
$O(\langle |F_{\alpha}|^2\rangle)=O(\kappa_0^{-2} m_{3/2}^2)$.
In most of the previous studies, motivated by
the fully renormalized vanishing cosmological constant,
the tree level expectation value of the hidden sector
scalar potential, $V_0\equiv \langle V_h\rangle_{\rm tree}$,
was assumed to
be zero or at least $\kappa_0^2V_0\ll m_{3/2}^2$.
However, as we have already anticipated in Eq. (3),
for a large number of chiral multiplets, $N=O(8\pi^2)$,
and the choice of the cutoff, $\kappa_0\Lambda=O(1)$,
quadratically divergent  quantum
correction
to the vacuum energy density becomes  of order $\kappa_0^{-2}m_{3/2}^2$.
This implies
$\kappa_0^2 V_0=O( m_{3/2}^2)$,
and then
one can not ignore\footnote{
Recently Brignole, Ibanez and Munoz$\cite{ibanez}$
also discussed the dependence
of soft parameters on $\langle\kappa_0^2V_h\rangle$ in the context of
string-inspired supergravity models. However, they did not
provide any rationale for $\langle\kappa_0^2 V_h\rangle $
to be of order $m_{3/2}^2$.} the $\langle
V_h\rangle$-dependent part of $m_0^2$ and $B$ in Eq. (11).

At one loop, the vacuum energy density receives a contribution
of the form$\cite{zumino}$
\be
V_{\rm eff}-(V_{\rm eff})_{\rm tree}\, =\,
\int {d^4p\over (2\pi)^4} \, {\rm Str} \,
[\ln (p^2+{\cal M}^2)]\, \simeq \,
\frac{1}{8\pi^2}{\rm Str} \,  ({\cal M}^2) \, \Lambda^2.
\ee
Here $(V_{\rm eff})_{\rm tree}$ is
the tree level value of $V_{\rm eff}$, thus it equals
$\langle V_h\rangle _{\rm tree}$.
For our simple model, we have
\be
{\rm Str} \, ({\cal M}^2) \,\simeq\, (Nm_0^2-\tilde{N}\tilde{m}^2)
+({\rm hidden\ sector\
contribution})\, \equiv \,
N_{\rm eff}m_0^2,
\ee
where $N$ is the number of observable chiral multiplets,
$\tilde{N}$ is the number of gauginos which
are assumed to have a common mass $\tilde{m}$, and
we neglected the masses of matter fermions and gauge
bosons. For the contribution from hidden sector,
the gravity sector
gives a negative contribution  ($=-4m_{3/2}^2$) while
hidden chiral multiplets give a positive contribution
proportional to the number of hidden multiplets.
Using Eq. (11), one easily finds
\def\l{\langle}
\def\r{\rangle}
\be
V_{\rm eff}-(V_{\rm eff})_{\rm tree} \, \simeq \, \frac{N_{\rm eff}}{8\pi^2}
m_0^2\Lambda^2
\, \simeq  \, \frac{N_{\rm eff}}{8\pi^2}\Lambda^2(m_{3/2}^2+\kappa_0^2 \l
V_0\r ).
\ee
Then the requirement of  $V_{\rm eff}=0$ leads to
\be
\kappa_0^2 \l V_h\r _{\rm tree} \, \simeq \, -\frac{\epsilon}{1+\epsilon}
m_{3/2}^2,
\ee
where
\be
\epsilon=\frac{N_{\rm eff}}{8\pi^2}(\kappa_0\Lambda)^2.
\ee

Unlike the cosmological constant, $\langle V_h\rangle$ receives
contributions only
from the hidden field fluctuations.  For $\kappa_0\Lambda=
g_{\rm GUT}$, if (i) interactions among hidden sector fields are weak
and (ii) the number of hidden multiplets which contribute to
supersymmetry breaking is $O(1)$, we have
\be
\delta \langle V_h\rangle=O\left({\alpha_{\rm GUT}\over \pi}|F_\alpha|
^2\right)=O\left({\alpha_{\rm GUT}\over \pi}\kappa_0^{-2}m^2_{3/2}\right)
\ee
where $\delta\langle V_h\rangle=\langle V_h\rangle
-\langle V_h\rangle_{\rm tree}$, and $F_\alpha$ is the $F$--term of
$h_{\alpha}$,$F_\alpha=D_\alpha W_h \exp (\kappa_0^2 h_\beta h^*_\beta
/2)=O(\kappa_0^{-2}m^2_{3/2})$. Note that conditions (i) and (ii) are
satisfied by many simple hidden sector models.

Thus we expect that $V_{\rm eff}\ne \langle V_h\rangle$ and the
difference is $O(\kappa_0^{-2}m^2_{3/2})$.

Since the gaugino mass contribution
is expected to be significantly smaller than the
chiral matter contribution, it is quite conceivable
that
$N_{\rm eff}$ is {\it positive} and of $O(8\pi^2)$.
Note that $N_{\rm eff}$ receives a contribution
from all chiral multiplets with masses far below
$M_P$, particularly from those in the minimal supersymmetric standard
model with 49 chiral multiplets.
Then for $\kappa_0\Lambda=g_{\rm GUT}$,
which is motivated by string theory, $\epsilon$
is essentially of order unity.
This implies that the soft scalar mass $m_0^2$ and the
part of $B$ associated with the K\"{a}hler potential term
$(\xi H_1H_2+{\rm h.c.})$ can be
significantly
affected by the contribution from $\kappa_0^2\l V_h\r$.
This also means that
to have a fully renormalized vanishing cosmological constant,
the hidden sector scalar potential $V_h$
is required to have a  {\it negative} expectation value
$\langle V_h\rangle$  of $O(\kappa_0^{-2} m_{3/2}^2)$ (see Eq. (19).).
In this regard, we note that,
in the `racetrack' model$\cite{racetrack}$ of gaugino condensations
for SUSY breaking in string theory,
the dilaton potential
appears often to have a negative minimum value.
Our discussion here indicates that a negative minimum
of the dilaton  potential
is not a problem, but is a rather
desirable feature for the fully
renormalized cosmological constant to vanish.
In passing, we note that the cosmological constant cannot be
solved purely from the high energy physics alone; one
has to deal with the quadratic
divergence of the minimal supersymmetric standard model.  If
the tree level cosmological constant is zero, then the full
cosmological constant is not zero.

One can mimick the above phenomenon in a simple quantum mechanical
toy model.  Let us introduce ($N+1$) dynamical variables,
$\{p_0, q_0\}$ and $\{p_i, q_i\}\ (i=1,\cdots, N)$,
\be
H= {1\over 2}p_0^2+(E_0+{1\over 2}\omega_0^2q_0^2)
+\sum^N_{i=1}\left[{1\over 2}p_i^2+{1\over 2}\left\{ \omega^2
+\lambda\omega_0 (E_0+{\omega_0^2q_0^2\over 2})
\right\} q_i^2\right]
\ee
where $E_0, \omega_0, \omega$ and $\lambda$ are parameters.
This toy model is written such that the following correspondence
between the previous supergravity model and the toy model
makes a sense:

Hidden sector potential energy operator, $V_h\equiv E_0
+{1\over 2}\omega_0^2q_0^2$

Mass (or frequency) of observable variables, $m^2\equiv
\omega^2+\lambda\omega_0 \l V_h\r$

The hidden sector ground state energy, $\tilde V\equiv \l {1
\over 2}p_0^2+V_h\r$

The true ground state energy, $V_{\rm eff} \leftrightarrow
\l H\r$

In the toy model, the classical ground state energy is given by
the condition, $p_0=p_i=q_0=q_i=0$.  Namely, the tree level values
are
\be
\l V_h\r _{\rm tree}=E_0,\ \ \ \ (V_{\rm eff})_{\rm tree}=E_0.
\ee
In this toy model also, the tree level values of $\l V_h\r$ and
$V_{\rm eff}$ are the same.  As before, we want to satisfy
$V_{\rm eff}=0$, not $(V_{\rm eff})_{\rm tree}=0$.

If we require $\l V_h\r_{\rm tree}=(V_{\rm eff})_{\rm tree}=0$,
then
\be
(m^2)_{\rm tree}=\omega^2.
\ee
Now we can include the quantum effects.  For ease of
demonstration, let us assume
\be
\omega_0\simeq \omega,\ \ \ \lambda\simeq {1\over N}
\simeq {\omega\over E_0}\simeq {1\over 8\pi^2}.
\ee
Then
\be
\l H_{h}\r =\l {1\over 2}p_0^2+E_0+{1\over 2}\omega_0^2q_0^2\r
=\l E_0+{\omega_0\over 2}\r = E_0\left(1+O({1\over 8\pi^2})\right)
\ee
and
\be
\l V_h\r =\l E_0+{1\over 2}\omega_0^2q_0^2\r =\l E_0+{\omega_0\over 4}\r
=E_0\left(1+O({1\over 8\pi^2})\right).
\ee
After performing the Gaussian integration with $\{p_0,q_0\}$,
we obtain the low energy Hamiltonian,
\be
H_{\phi}\ =\ \left(\sum^N_{i=1}{1\over 2}p_i^2\right)
+\left[E_0+{1\over 2}\omega_0+{1\over 2}(\omega^2+\lambda\omega_0
E_0)\sum^N_{i=1}q_i^2\right]+O(\lambda^2)
\ee
from which we can read the effective potential
\be
V_{\rm eff}\ =\ E_0+{1\over 2}N\sqrt{\omega^2+\lambda\omega_0 E_0}
+O(\omega).
\ee
Thus, we obtain
\be
V_{\rm eff}-\l V_h\r ={1\over 2}N\sqrt{\omega^2+\lambda \omega_0 E_0}
+O(\omega)
\ee
which has our previous
form of $(1/8\pi^2)\times $(number of chiral fields).

In the above discussion, we considered only
the one loop contribution (Fig. 1)
to the vacuum energy density, which is
of order $\kappa_0^{-2} m_{3/2}^2$ due to a
large value of $N_{\rm eff}$ compensating over
the loop
suppression factor $1/8\pi^2$.
Clearly this contribution persists even when all
interactions
(of course except for the kinetic and mass part)
 in the supergravity action are turned off.
Inclusion of interactions gives rise to additional quadratic
divergences, but at higher loop order.
Contrary to the one loop contribution,
these higher loop effects do {\it not} contain any
additional large factor
which may compensate over the additional loop suppression
factor.  Obviously
two loop diagrams involving gauge interactions give
a contribution smaller than the one loop
effect by the small factor $\alpha_{\rm GUT}/2\pi$.
Note that quadratically divergent effects are dominated
by the contribution from fluctuations at the cutoff scale
where all interactions are presumed to be perturbative.
\vskip 5.5cm
\centerline{Fig. 1. One loop contribution to the cosmological
constant.}

One can consider other types of two loop diagrams
involving trilinear and/or Yukawa scalar interactions.
(See  Figs. 2 and 3.)  The contribution of Fig. 2 is roughly
\be
\frac{1}{(4\pi^2)^2}\sum_{ijk}A_{ijk}A^*_{ijk}
\Lambda^2,
\ee
while that of  Fig. 3 is
\be
\frac{1}{(4\pi^2)^2}\sum_{ijk}\lambda_{ijk}\lambda^*_{ijk}
m_{3/2}^2\Lambda^2.
\ee
Usually $A_{ijk}$ is of order $m_{3/2}\lambda_{ijk}$,
and then clearly the above two loop effects are negligible
compared to the one loop effect of Eq. (18).
\vskip 6cm
\centerline{Fig. 2. The $A$ term contribution at two loop level.}
\vskip 9cm
\centerline{Fig. 3.Two loop contribution from Yukawa interaction.}
Since the quantum corrections to the cosmological constant of order
$\kappa_0^{-2}m_{3/2}^2\gg m_W^4$ are added to the tree level value,
one might wonder whether
the vacuum structure of the observable sector fields $\phi_i$
is changed.   Our results of Eqs. (11) and (18)
explicitly show that
soft masses of $\phi_i$
 at the cutoff scale are  still {\it positive}
even after including the contribution
from a {\it negative} $\langle V_h\rangle$.
It is also easy to see that ${\rm Str} \, ({\cal M}^2)$
is {\it independent} of $\phi_i$ and
thus the vacuum structure of $\phi_i$ is untouched
by the quadratically divergent corrections discussed above.
To see it more explicitly, we can consider a simple example
with the superpotential $W=\lambda \phi^3+m\phi^2$ and the soft
term ${\cal L}_{\rm soft}=m_0^2\phi\phi^*$. We then have, for
an arbitrary value of $\phi$,
$$
{\rm Str} \, ({\cal M}^2)=m_0^2+ |6\lambda\phi+2m|^2-|6\lambda\phi+2m|^2,
$$
which is $\phi$-independent.
Clearly this is a simple consequence of the nonrenormalization
theorem for the effective potential in the presence of softly broken
SUSY, particularly  the absence of a field-dependent quadratic
divergence.

So far we have shown that the quantum corrections to
the cosmological constant in supergravity models can be of
order $\kappa_0^{-2}m_{3/2}^2$ and
most likely will be positive.
This is for the reasonable  choice of the cutoff scale,
$\kappa_0\Lambda=O(g_{\rm GUT})$, and mainly due to
a large number of chiral multiplets which conpensates over
the loop suppression factor.
This implies that conventional studies of
supergravity phenomenology based on the input
$\kappa_0^{2}\langle V_h\rangle \ll m_{3/2}^2$ should be modified.
This also implies that it is rather desirable,
for the fully renormalized vanishing cosmological constant to
vanish, to have a negative tree level cosmological constant
of order $\kappa_0^{-2}m_{3/2}^2$.

In our scheme, the squark
masses $m_i$ can be expressed as
\be
{m_i^2\over m^2_{3/2}}\ =\ 3C^2\left\{1+{1\over 3}N_i(T,T^*)\cos^2
\theta\right\}-2
\ee
where $\cos\theta$ signifies the relative importance in supersymmetry
breaking by moduli and dilaton superfields (e.g. $\cos\theta\sim 0$
corresponds to the dilaton dominated supersymmetry breaking and vice
versa) and $N_i$ is related
to the curvature of the K$\ddot {\rm a}$hler manifold.  For (2,2)
Calabi-Yau spaces, $N_i\rightarrow -1$ as the moduli field $T
\rightarrow \infty$. For orbifolds, $N_i$ corresponds to
modular weights of charged fields which are normally negative
integer numbers\cite{luest}.  Not to break $SU(3)_c\times
U(1)_{\rm em}$ gauge symmetry, we must require
\be
C^2>{2\over 3+N_i(T,T^*)\cos^2\theta}.
\ee
Because $\langle V_h\rangle$ most probably will be negative
as discussed in Eq. (29), we would have $C^2<1$
(viz. $C^2=1+(8\pi\langle V_h\rangle/3m^2_{3/2}M_{Pl}^2)$).
At present, it is difficult to give any precise value for $C^2$
in view of the lack of information on the mass spectra of
superpartners, and hence the phenomenological considerations studied
for the case $C^2=1$ should serve only as
a guideline \cite{barbieri}.
For scalar masses, this contribution can shift them by $O(10)$ \%,
depending on the magnitude of $\epsilon$.

In conclusion, I discussed the quantum correction to the
cosmological constant in supergravity and its effects on the
soft parameters.  Even though the $N=1$ supergravity is not
renormalizable, this calculation seems to be reliable if
the theory is replaced by a well behaved one above the cutoff
scale.  Without supersymmetry, this loop calculation has the
quartic divergence and hence the discussion must be postponed
until a renormalizable theory of gravity is found.

\acknowledgments
I thank Professors H.-Y. Guo and S. Chang for the invitation to
this Workshop and their kind hospitality extended to me during
my stay in Beijing.
I have benefitted from discussions with Kiwoon Choi and
Hans Peter Nilles.  This work is supported in part by Korea Science
and Engineering Foundation through Center for Theoretical Physics at
Seoul National University,  KOSEF--DFG Collaboration
Program, and the Ministry of Education of The Republic of
Korea through the Basic Science Research Institute, Contract No.
BSRI-94-2418.

\end{document}